\documentclass[twocolumn,showpacs,preprintnumbers,amsmath,amssymb]{revtex4-1}

\usepackage{graphicx}
\usepackage{bm}

\begin{document}

\title{The quasi time crystal}

\author{Xintian Wu}
\email{Corresponding author: wuxt@bnu.edu.cn}
\affiliation{Department of Physics, Beijing Normal University,
Beijing, 100875, China}

\date{\today}

\begin{abstract}
We discuss the possibility of making a quasi time crystal. A simple two-state model is studied to clarify our definition. In a superposition of the ground state and the excited state and the probability of observation varies periodically in time during the lifetime of the excited state. 
The quasi time crystal is also discussed around the first order quantum phase transition, which is characterized by the degeneracy and crossing of the two lowest-energy states in the infinite-volume limit.  Our results have broad validity. As an example, the one-dimensional transverse field Ising model with surface fields is shown to have similar behavior. The oscillating magnetization profile is solved exactly.
\end{abstract}

\maketitle

Ten years ago, Wilczek proposed the concept of time crystals, which spontaneously break the continuous time translation symmetry \cite{wilzeck,wilzeck1}. The proposal stimulated intense debates and many new ideas. The original concept was that the continuous time translation symmetry is spontaneously broken in the ground state or equilibrium in analogy with ordinary crystals that break the continuous spatial translation symmetry. However this idea and the proposals to realize it experimentally provoked many questions \cite{li,bruno,bruno1,watanabe}. Watanabe and Oshikawa presented a definition of time crystals based on the time-dependent correlation functions of the order parameter \cite{watanabe}. They proved a no-go theorem that ruled out the possibility of time crystals defined as such, in the ground state or in the canonical ensemble of a general Hamiltonian, which consists of not-too-long-range interactions.

Later, the time crystal was proposed in the driven systems with Hamiltonian being periodic in time \cite{sacha,nayak,khemani,yao}. The theoretical and experimental development along these lines was quite successful \cite{zhang,kyprianidis,choi,sullivan,rovny,pal,smits,autti,randal}. However, in such systems the discrete, rather than the continuous, time translation symmetry is broken because the Hamiltonian is periodic in time.

In this note, I try to view the issue from a different perspective and make a time crystal that breaks the continuous translation symmetry. In the first paper in which Wilczek mentioned the time crystal, he wrote that for an operator without internal time dependence
\begin{equation}
\langle \Psi|\dot{O}|\Psi \rangle=\langle \Psi| [H,O] |\Psi \rangle=0
\end{equation}
if $\Psi=\Psi_E$ is an eigenstate of $H$. This seems to preclude the possibility of an order parameter that could indicate the spontaneous breaking of infinitesimal time-translation symmetry.

If $\Psi$ is not an eigenstate, for example, but a superposition of an excited state and the ground state, it is possible to break the continuous time-translation symmetry. In this case, the observable expectation can change periodically in time. Of course, the excited state will be dissipated due to the spontaneous transition to the ground state. However, if the lifetime of the excited state is long enough, the periodic change of an observable expectation can occur.
To clarify my idea, I use a simple model with two states
\begin{equation}
H=-\delta {\Big (}|1\rangle \langle 2|+|2\rangle \langle 1|{\Big )}
\label{eq:ham}
\end{equation}
where $|1 \rangle,|2 \rangle$ are two base states and $\delta >0$. This model can describe many different systems. For convenience, I consider it to be a double-well ion trap. Then, $|1 \rangle,|2 \rangle$ refer to the first and second wells, respectively, and $\delta$ is the hopping term between the two wells. The site energies are set to zero. 
It has two eigenstates $ |\pm \rangle=\frac{1}{\sqrt{2}}{\Big (}|1 \rangle \pm|2 \rangle{\Big )}$ with eigenvalues $ E_{\pm}=\mp \delta$. If the initial state is set to\begin{equation}
|\varphi(t=0) \rangle=|1 \rangle =\frac{1}{\sqrt{2}}{\Big (}|+\rangle +|- \rangle{\Big )}.
\label{eq:super}
\end{equation}
it is the superposition of the ground state and the excited state. Then, at later $t$, the state vector is given by
\begin{equation}
|\varphi(t)\rangle=\frac{1}{\sqrt{2}}{\Big (}|+\rangle e^{-iE_+t}+|-\rangle e^{-iE_-t}{\Big )}.
\end{equation}
At time $t$, the probability of an ion being in the first well is given by $|\langle 1|\varphi(t)\rangle|^2=\cos^2 2 \delta t $, and at the second well is,
$ |\langle 2|\varphi(t)\rangle|^2=\sin^2 2 \delta t$. We assume that the well sizes are much smaller than the distance between the two wells. If the positions of the first and second wells are given by ${\textbf r}_1,{\textbf r}_2$, then the ion position expectation is given by
\begin{equation}
\langle \varphi(t)|{\textbf r}|\varphi(t)\rangle=\frac{1}{2}{\textbf r}_1 \cos^2 2 \delta t+\frac{1}{2}{\textbf r}_2 \sin^2 2 \delta t.
\end{equation}
and it varies with time and breaks the continuous time-translation symmetry. The change in the position expectation over time is due to the interference between the two eigenstates.

Of course, this state cannot exist perpetually because the excited state will transition to the ground state spontaneously. However, if the lifetime of the excited state is long enough, periodic changes in the ion position can be observed. This phenomenon can be called a quasi-time crystal, similar to a quasi-particle whose lifetime is finite.

The initial state Eq. (\ref{eq:super}) can be realized through quantum quenching \cite{cardy}. Set the ion trap Hamiltonian to be $H_0=-A |1\rangle \langle 1|-\delta (|1\rangle \langle 2|+|2\rangle \langle 1|)$ and $A\gg \delta$ and prepare the system in the ground state of $H_0$, which is approximately $|1 \rangle$. At time $t=0$, varying the parameter $A$ to be $0$, the Hamiltonian is changed to be $H$ defined in Eq. (\ref{eq:ham}). This variation, or quenching, is supposed to be carried out over a time scale much less than $\delta^{-1}$. For times $t>0$, the system evolves unitarily according to the dynamics given by $H$.

The above phenomena should occur in many body systems. The first-order quantum phase transition in many body systems is caused by the crossing of the two lowest-energy states in the infinite-volume limit \cite{vicariprl}. This is very similar to the two-state model above. However, the phenomena in many body systems are more complicated, and usually there is no exact solution. Here, I take the one-dimensional transverse field Ising model as an example. One advantage is that this case can be solved exactly.

Consider the one-dimensional transverse field Ising model with surface fields. Its Hamiltonian is given by
\begin{equation}
H=H_0-h_L \sigma^{(1)}_1-h_R \sigma_N^{(1)}
\label{eq:hami}
\end{equation}
where 
\begin{equation}
H_0=-\sum_{n=1}^{N-1} \sigma^{(1)}_n \sigma^{(1)}_{n+1}-g\sum_{n=1}^{N}\sigma^{(3)}_n,
\end{equation}
$\sigma_i^{(1)},\sigma_i^{(3)}$ are Pauli matrices; $N$ is the length of the Ising chain; and $h_L,h_R$ are the left and right boundary longitudinal fields, respectively. 

\begin{figure}
\includegraphics[width=0.5\textwidth]{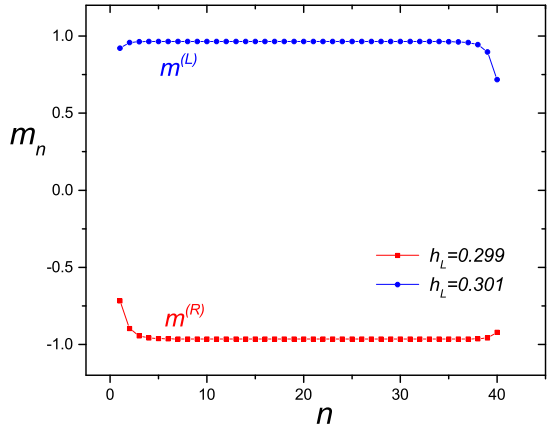}
\caption{The magnetization profiles around the first-order phase transition at $h_L=|h_R|$ with $h_R=-0.3$ and $g=0.5$. The length of the chain is $N=40$. The numerical calculation is carried out with double precision.}
\end{figure}

For $g<1$, the Ising chain is in the ordered phase. It is shown that for $h_L=-h_R$ and $|h_L|,|h_R|<\sqrt{1-g}$, there is a first-order transition \cite{wu}. For $h_L-|h_R| \gg \kappa$, where $\kappa \propto [(1-h_R^2)/g]^{-N}$ decreases exponentially with the system size $N$, the left surface field dominates, and the system is in the ``positive" phase. In contrast, for $|h_R|-h_L \gg \kappa$, the right surface field dominates, and the system is in the ``negative" phase. 

We show the dramatic change in the magnetization profile with $g=0.5,h_R=-0.3,N=40$ in Fig. 1, where $m_n$ is the expectation of $\sigma_n^{(1)}$. For $h_L=0.299<|h_R|$, the right boundary field dominates, and the magnetization of most spins is negative. We call this state the ``negative" phase. The corresponding magnetization profile is presented by the line scatter in red, which is denoted by $m^{(R)}$. For $h_L=0.301>|h_R|$, the left boundary field dominates, and the magnetization is positive for the majority of the spins. We call this state the ``Positive" phase. The corresponding magnetization profile is presented by the line scatter in blue, which is denoted by $m^{(L)}$.

If the system is in the ``positive" phase and varying $h_L$ to be $h_L=-h_R$ by quantum quench \cite{cardy}, for example, initially the system is at the ground state with $h_R=-0.3,h_L=0.301$ at $t=0$ and varying $h_L$ to be $h_L=-h_R=0.3$ by quantum quench, for times $t>0$ the magnetization profile evolves according to
\begin{equation}
m_n\approx \frac{1}{2}{\Big (}m_n^{(L)}+m_n^{(R)}{\Big )}+\frac{1}{2}{\Big (}m_n^{(L)}-m_n^{(R)}{\Big )}\cos \omega t.
\label{eq:magosc}
\end{equation}
where $\omega$ is the energy gap at $h_L=-h_R=0.3$. That is, the magnetization profile oscillates back and forth between $m^{(L)}$ and $m^{(R)}$ periodically.

This state is also a superposition of the ground state and the first excited state, which is almost degenerated with the ground state. The oscillation of the order parameter is also caused by the interference between the two states. Again, this state will be dissipated. However, in the lifetime of the excited state, the oscillation of the order parameter can be observed. This is another example of the quasi-time crystal. 

Now, we give the proof. Following the well-known theories \cite{bariev,hinrichsen,bilstein,vicari}, we transform the diagonalization problem to an effective Hamiltonian by appending one additional spin to the left and right sides. The corresponding effective Hamiltonian is given by
\begin{equation}
H_e = H_0 -|h_L| \sigma^{(1)}_0 \sigma^{(1)}_1-|h_R| \sigma^{(1)}_N\sigma^{(1)}_{N+1}. 
\label{eq:hamiltonian}
\end{equation}
Because $\sigma_0^{(1)}, \sigma_{N+1}^{(1)}$ are free from the transverse field, both $\sigma_0^{(1)},\sigma_{N+1}^{(1)}$ commute with the Hamiltonian. Hence, they can be diagonalized simultaneously. 

The Hilbert space can be divided into four sectors, which we label $(1,1),(1,-1),(-1,1),(-1,-1)$, where $(s_0,s_{N+1})$ are eigenvalues of $\sigma^{(1)}_0$ and $\sigma^{(1)}_{N+1}$. The restriction of $H_e$ to the four sectors gives rise to the Hamiltonian $H$ with four cases of different signs of $h_L,h_R$ \cite{vicari}. For example, the restriction of $H_e$ to sector $(1,-1)$ gives rise to the Hamiltonian $H$ with $h_L>0,~h_R<0$. 

To compute the spectrum of the Hamiltonian, Eq. (\ref{eq:hamiltonian}), we perform the Jordan-Wigner transformation and define fermionic operators
$ c_n^{\dagger}=(-1)^{n}\prod_{l=0}^{n-1}\sigma^{(3)}_l\sigma_n^+$, where $\sigma^{\pm}=(\sigma^{(1)}\pm i \sigma^{(2)})/2$. The Hamiltonian becomes
\begin{equation}
H_e=-gN+\sum_{n,l=0}^{N+1}{\Big (}c_n^{\dagger}{\bf A}_{n,l}c_l+\frac{1}{2}c_n^{\dagger}{\bf B}_{n,l}c_{l}^{\dagger}-\frac{1}{2}c_n {\bf B}_{n,l}c_l{\Big )}
\end{equation}
where ${\bf A}$ and ${\bf B}$ are symmetric and antisymmetric matrices, respectively.

We perform a Bogoliubov transformation by introducing new canonical fermionic variables \cite{lieb}
\begin{equation}
\eta_k=\sum_{n=0}^{N+1} {\Big (}\frac{\phi_{k,n}+\psi_{k,n}}{2}c_n+\frac{\phi_{k,n}-\psi_{k,n}}{2}c_n^{\dagger}{\Big )},
\end{equation}
where eigenvector $\psi_k,\phi_k$ are given by 
\begin{equation}
{\bf C}\psi_k=\varepsilon_k^2 \psi_k, ~~~\phi_k=({\bf A}-{\bf B})\psi_k /\varepsilon_k,
\label{eq:AB}
\end{equation}
where ${\bf C}\equiv ({\bf A}+{\bf B})({\bf A}-{\bf B})$, and $\varepsilon_k$ is the eigenenergy of the fermion $\eta_k$. Therefore, the problem is transformed to a free fermion problem. If ${\bf C}$ is diagonalized, all the physical quantities can be calculated \cite{lieb}. 

It should be emphasized that the ground state and the first excited state of $H$ with $h_L>0,h_R<0$ are the first and second excited state of $H_e$ respectively.
They belong to sector $(1,-1)$ and are given by
\begin{equation}
|\Psi_1\rangle=\eta_1^{\dagger} |\Psi_0\rangle, \hskip 0.5cm |\Psi_2\rangle=\eta_2^{\dagger} |\Psi_0 \rangle.
\label{eq:ground}
\end{equation}
where $| \Psi_0 \rangle $ is the ground state of $H_e$. 

We set $h_R<0$ and $|h_R|<\sqrt{1-g}$ and vary $h_L$ around the phase transition point, i.e., $|h_L+h_R|\ll 1$. In this case, there exist two localized state eigenvectors of matrix ${\bf C}$, whose details can be seen in section I of the supplementary material. For an Ising chain with size $N \gg 1$, these eigenvectors are given by
\begin{equation}
\psi=u \psi^{(L)}+ v\psi^{(R)}
\label{eq:mix}
\end{equation}
where\begin{eqnarray}
\psi^{(R)}_0 & = &\frac{\beta x^{-N}}{h};~~ \psi^{(R)}_n= (-1)^j\beta x^{n-N}, ~~for ~~ n\ge 1 \nonumber \\ 
\psi^{(L)}_0 & = & \frac{\alpha}{h}; ~~\psi^{(L)}_n= (-1)^n\alpha x^{-n}, ~~ for ~~n\ge 1 
\label{eq:mix}
\end{eqnarray}
with $ \alpha=h\sqrt{x^2-1}/ \sqrt{x^2+h^2-1}$, $\beta=\sqrt{x^2-1}/x$. For $x>1$, the wavefunction $\psi_L$ is localized at the left end and $\psi_R$ at the right end.The eigenvalues of these eigenvectors satisfy\begin{equation}
\varepsilon^2=4{\Big [}1+g^2-g(x+x^{-1}){\Big ]}.
\end{equation}

For $|h_L+h_R|\ll 1$, the parameter $x$ satisfies the following equations:
\begin{eqnarray}
(x-x_L)u-\delta_R v & =0 \nonumber \\
-\delta_R u+(x-x_R)v & =0
\end{eqnarray}
where $x_L=(1-h_L^2)/g$, $x_R=(1-h_R^2)/g$, and $\delta_R \equiv \alpha(x_R-x_R^{-1})x_R^{-N}/\beta$. In the thermodynamic limit $N\rightarrow \infty$, $\delta_R=0$. Then, we obtain two simple solutions $x=x_L$ and $x=x_R$. They correspond to the eigenvector $\psi_L$ localized at the left end and $\psi_R$ at the right end, respectively. Their eigenvalues are given by $\varepsilon_{L,R}=2\sqrt{1+g^2-g(x_{L,R}+x_{L,R}^{-1})}$. Fixing $h_R$, $x_R$ and $\varepsilon_R$ are fixed. Varying $h_L$, $x_L$ and $\varepsilon_R$ are changed. At the phase transition point $h_L=-h_R$, the eigenenvalues of the two eigenvectors $\psi_L$ and $\psi_R$ degenerate. At this level crossing point, a first-order phase transition occurs \cite{sachdev}.

For a lattice with finite size $N\gg 1$, at the transition point $h_L=-h_R$, the degeneracy of the two localized eigenvectors is lifted by the nonzero $\delta_R$. Then, there are two real roots for $x$:
\begin{equation}
x_1=x_R+\delta_R,~~x_2=x_R - \delta_R
\label{eq:roots}
\end{equation}

The corresponding eigenvectors are given by
\begin{equation}
\psi_1=\frac{1}{\sqrt{2}}{\Big (}\psi^{(L)}+\psi^{(R)}{\Big )}, ~~\psi_2=\frac{1}{\sqrt{2}}{\Big (}\psi^{(L)}-\psi^{(R)}{\Big )}
\end{equation}
and their eigenvalues are given by\begin{equation}
 \varepsilon_{1,2}=2\sqrt{1+g^2-g(x_{1,2}+x^{-1}_{1,2})}.
\end{equation}
Since $\delta_R \ll 1$, these two eigenstates are nearly degenerated.

Suppose the initial state is given by
\begin{equation}
|\varphi (t=0)\rangle =\frac{1}{\sqrt{2}}{\Big (}\eta_1^{\dagger}+\eta_2^{\dagger}{\Big )}|\Psi_0\rangle,
\label{eq:initial-m}
\end{equation}
which is the superposition of the ground state and the first excited state. At later $t$, the state vector evolves unitarily according to the dynamics given by $H$ with $h_L=-h_R$ and is given by\begin{equation}
|\varphi (t)\rangle =\frac{1}{\sqrt{2}}{\Big (}\eta_1^{\dagger}e^{-i\varepsilon_1 t}+\eta_2^{\dagger}e^{-i\varepsilon_2 t}{\Big )}|\Psi_0\rangle
\end{equation}

The magnetization of the $n$th spin in this state should be given by
\begin{equation}
m_n=\langle \varphi (t)| \sigma_n^{(1)}|\varphi (t)\rangle=\frac{1}{s_0}\langle \varphi (t)|\sigma_0^{(1)} \sigma_n^{(1)}|\varphi (t)\rangle,
\label{eq:magnetization1}
\end{equation}
where $s_0$ is the eigenvalue of $\sigma_0^{(1)}$ because $\sigma_0^{(1)}$ commutes with $H_e$. Here, we consider $h_L>0$; then, it has $s_0=1$. Through the calculation of the above correlation, we can obtain the magnetization.

To calculate the magnetization, we define the operators\begin{equation}
\eta_{L} = \frac{1}{\sqrt{2}}{\Big (}\eta_1+\eta_2{\Big )},~~\eta_{R} = \frac{1}{\sqrt{2}}{\Big (}\eta_1-\eta_2{\Big )}
\end{equation}
and vectors
\begin{equation}
|\Psi^{(L)}\rangle =\eta_{R}^{\dagger}|\Psi_0 \rangle, ~~|\Psi^{(R)}\rangle =\eta_L^{\dagger}|\Psi_0 \rangle.
\end{equation}
We find that\begin{equation}
\langle \Psi^{(L)}|\sigma_0^{(1)} \sigma_j^{(1)}|\Psi^{(R)}\rangle \approx \langle \Psi^{(R)}|\sigma_0^{(1)} \sigma_j^{(1)}|\Psi^{(L)}\rangle \approx 0. 
\label{eq:zero}
\end{equation}
After some calculation, we obtain Eq. (\ref{eq:magosc}) with\begin{equation}
\omega =\varepsilon_2-\varepsilon_1
\end{equation}
which is the energy gap, and
\begin{eqnarray}
m_n^{(L)} & = & \langle \Psi^{(L)}|\sigma_0^{(1)} \sigma_n^{(1)}|\Psi^{(L)}\rangle \nonumber \\
m_n^{(R)} & = & \langle \Psi^{(R)}|\sigma_0^{(1)} \sigma_n^{(1)}|\Psi^{(R)}\rangle.
\label{eq:nonzero}
\end{eqnarray}
The magnetization profile $m_n^{(R)}$ is in the ``negative" phase for $h_L<|h_R|$, where the right surface field dominates. In this case, $\psi^{(L)}$ is the ground state eigenvector of ${\bf C}$, and the many body ground state is given by $\eta_L^{\dagger}|\Psi_0\rangle$. Therefore, we obtain the first one in the above equations. There is a similar argument for the second equation. The proof of Eq. (\ref{eq:zero}) and the computational procedure for Eq. (\ref{eq:nonzero}) are discussed in the supplementary material \cite{sup}.

To realize the initial state, Eq. (\ref{eq:initial-m}), one can prepare the system in the ground state of $H$ with $h_L>|h_R|$ and $1\gg h_L-|h_R| \gg g\delta_R/2h_R$. Generally, the ground state of $H_e$ depends on $h_L$. However, if $ h_L-|h_R|\ll 1$ is satisfied, the variation in the ground state of $H_e$ is negligible. In the regime we are concerned, the ground state of $H_e$ is approximately $|\Psi_0 \rangle$ with $h_L=-h_R$. Consequently, the ground state of $H$ in Eq. (\ref{eq:hami}) is given by $\eta_L^{\dagger}|\Psi_0 \rangle=\frac{1}{\sqrt{2}}{\Big (}\eta_1^{\dagger}+\eta_2^{\dagger}{\Big )}|\Psi_0\rangle$ approximtely for $h_L>-h_R$. At time $t=0$, varying the parameter $h_L$ to be $h_L=-h_R$, the Hamiltonian is changed to be $H$ defined in Eq. (\ref{eq:hami}) with $h_L=-h_R$. This variation, or quenching, is supposed to be carried out over a time scale much smaller than $\omega^{-1}$. For times $t>0$, the system evolves unitarily according to the dynamics given by $H$ with $h_L=-h_R$. 

Our results have broad validity and, in particular, apply to any first-order quantum transition characterized by the degeneracy and crossing of the two lowest-energy states in the infinite-volume limit. Almost all of the systems described by the Landau-Ginzburg Hamitionian with a coupling term between the order parameter and the external field have a first-order phase transition in the ordered phase. The quantum versions of such systems should be characterized by the degeneracy and crossing of the two lowest-energy states in the infinite-volume limit \cite{vicariprl}. Therefore, a huge class of systems can be used to make the quasi-time crystal. 

To observe this kind of phenomenon requires a long lifetime of the excited state. It is well known that the lifetime of the excited state is the inverse of the Einstein probability coefficient for spontaneous transition to the lower state \cite{einstein,shordinger,dirac}. In general, electromagnetic radiation has a narrow linewidth. This means a long enough lifetime during which there are a huge number of cycles. Therefore, we can expect the lifetimes of the two examples discussed above to be long enough to observe the oscillation.
This kind of state has merit. The breaking of time-translation symmetry is spontaneous with respect to the continuous time-translation symmetry. Its disadvantage is that the state is transient. However, the transient state is not insignificant. Whether the lifetime is long or short depends on the timescale of concern. It is well known that for the phase transition, the symmetry is broken as the system size approaches infinity. For finite-size systems, there is no symmetry breaking in the absolute sense. Take the ferromagnet as an example. Below $T_c$ at $h=0$, the up and down macrostates are separated by a free energy barrier of height order $N^{1/2}$, giving a characteristic time $\tau_{flip}$ for reversal of order $e^{a N^{1/2}}$ \cite{palmer}. If the observational timescale $\tau_0$ is less than $\tau_{flip}$, the total magnetization is not zero, and the symmetry is effectively broken. Otherwise, the symmetry is not broken. 

It is plausible that the time-translation symmetry breaking can be observed not in a ground state associating that the pendulum swings only if it is not in equilibrium state and it stops if it is.

There is already a device to simulate the transverse field Ising model based on the technique of ultracold atoms \cite{islam,labuhn}. The suggested phenomena here should be tested in the near future.

It should be mentioned that Sacha et al. have studied discrete time quasicrystal \cite{sacha1}. The ``quasi" in theor work refers to the structure of  time crystal. In our work it refers to the finite timelife of time crystal.

The authors thank Wenan Guo and the SGI in the Department of Physics, Beijing Normal University, for providing computing time.

\end{document}